\documentclass[conference]{IEEEtran}
\usepackage{subfig}
\usepackage{listings}
\usepackage{cite}
\usepackage{color}
\usepackage{multirow}
\usepackage{url}
\usepackage[pdftex]{graphicx}
  \DeclareGraphicsExtensions{.pdf,.jpeg,.png}
\usepackage{array}
\usepackage{ctable}
\usepackage{amsmath}
\newtheorem{definition}{Definition}
\newcommand{\model}{ARUS}
\newcommand{\tool}{SALAD}
\newcommand{\hmm}{HAPI}

\newcommand{\code}[1]{{\footnotesize\textsf{#1}}}

\begin{document}
%
\title{Learning API Usages from Bytecode:\\A Statistical Approach}


\author{\IEEEauthorblockN{Tam The Nguyen,
Hung Viet Pham,
Phong Minh Vu,
Tung Thanh Nguyen}
\IEEEauthorblockA{Computer Science Department\\
Utah State University\\
\{tam.nguyen,hung.pham,phong.vu\}@aggiemail.usu.edu\\
tung.nguyen@usu.edu}}

\maketitle

\begin{abstract}
When developing mobile apps, programmers rely heavily on standard API frameworks and libraries. However, learning and using those APIs is often challenging due to the fast-changing nature of API frameworks for mobile systems, the complexity of API usages, the insufficiency of documentation, and the unavailability of source code examples. In this paper, we propose a novel approach to learn API usages from bytecode of Android mobile apps. Our core contributions include: i) {\model}, a graph-based representation of API usage scenarios; ii) {\hmm}, a statistical, generative model of API usages; and iii) three algorithms to extract {\model} from apps' bytecode, to train {\hmm} based on method call sequences extracted from {\model}, and to recommend method calls in code completion engines using the trained {\hmm}. Our empirical evaluation suggests that our approach can learn useful API usage models which can provide recommendations with higher levels of accuracy than the baseline n-gram model.
\end{abstract}



%
\IEEEpeerreviewmaketitle

\section{Introduction}

Millions of mobile apps (i.e. software applications for mobile devices like smartphones and tablets) have been developed and made available to mobile users. Due to the fierce competition, those mobile apps often have very short time-to-market and upgrade cycles, thus, requiring short development time. To address this requirement, programmers often rely heavily on API application frameworks and libraries such as Android and iOS frameworks, Java APIs, etc. For example, an Android app might make 8-42\% of its external dependencies to Android APIs and 7-68\% to Java APIs~\cite{hassan_cascon}. 

Learning and using APIs is often challenging due to several reasons. First, a framework often include large numbers of API elements (i.e. class or method). For example, Android application framework contains over 3,400 classes and 35,000 methods, clustered in more than 250 packages~\cite{wmapp}. Moreover, typical API usage scenarios often include several API elements and follow special rules, e.g. for pre- and post-conditions or for control and data flows~\cite{groum, grapacc, boa.icse2014}. Unfortunately, API documentation is often insufficient. For example, the Javadoc of a class often contains only descriptions of its methods and rarely has code examples on the usage of its objects and methods~\cite{sungkim.exoa}. Documentation and code examples for API usages involving several objects are often non-existed.

The situation is even more difficult for learning APIs of mobile frameworks. First, due to the fast development of mobile devices and the strong competition between vendors, those frameworks are often upgraded quickly and include very large changes. For example, 17 major versions of Android framework have been released within five years, making nearly 100,000 changes to its API methods~\cite{wmapp}. More severely, mobile apps are often closed-source, i.e. their source code is not publicly available. Thus, finding and learning code examples from existing mobile app projects would be difficult.

Due to those reasons, the current best API learning approach for mobile app programmers would be looking for online tutorials and discussions on developer forums like StackOverflow.com. However, this approach is rather ineffective and time-consuming, thus, could severely affect programming productivity. In addition, it is also error-prone because online code examples might not be up-to-date to the latest changes of the mobile API frameworks. When programmers do not learn from accurate materials, they could use the APIs improperly, causing errors and defects and thus, reducing the quality and the usability of their apps~\cite{wmapp,hassan_cascon}.

In this paper, we introduce a novel approach, {\tool}, to address those problems. Standing for {\it ``\underline{S}tatistical \underline{A}pproach for \underline{L}earning \underline{A}PIs from \underline{D}VM bytecode''}, our approach has three distinctive characteristics. First, to address the problem of insufficient documentation and source code examples, {\tool} learns API usages from \emph{bytecode} of Android mobile apps, which are publicly available in very large quantities (up to millions of apps). Second, {\tool} can learn complex API usages consisting large numbers of objects and methods. Finally, {\tool} can automatically generate recommendations for incomplete API usages, thus could reduce the chance of API usage errors and improve code quality.

Let us discuss {\tool} in more details. The most important component of {\tool} is {\hmm}, standing for {\it ``\underline{H}idden Markov Model of \underline{API} usages''}. A {\hmm} is an Hidden Markov Model (HMM)~\cite{hmm_intro} which models the usages of one or multiple API objects. As a statistical model, it has several internal states, each has a probability to be selected as the starting state. When a {\hmm} is in a state, it can generate a method call and change to another state with specified probabilities. We have designed an algorithm to train a {\hmm} (i.e. inferring its internal states and the associating probabilities) from a given collection of method call sequences involving the API objects it models. Another algorithm is designed to use a trained {\hmm} to compute the generating probabilities of several method call sequences to rank those sequences. This ranking result is used in the recommendation task of {\tool}. More details on {\hmm} and the training/ranking algorithms will be discussed in Section III and IV.

To extract and generate API method call sequences from bytecode for training {\hmm}, we develop another conceptual model named {\model}, standing for {\it ``\underline{A}bstract \underline{R}epresentation of  \underline{U}asge \underline{S}cenario"}. An ARUS is a graph-based model in which the nodes represent method invocations and data objects and the edges in the ARUS indicate control and data flows between those methods and objects. With that design, an ARUS can represent an usage scenario including many objects and methods and complex control and data flows between them. We have designed and implemented in {\tool} an algorithm to extract ARUS models directly from Android apps' bytecode. 
More details on {\model} and this algorithm will be discussed in Section III  and IV.

To evaluate the usefulness and effectiveness of {\tool}, we have conducted several experiments comparing the accuracy in recommending API usages of {\hmm} and the baseline $n$-gram model which is widely used in most prior studies. The experiments show two important results. First, {\hmm} can provide API usage recommendations with a high level of accuracy. For example, it can predict and recommend the next API call for a given method sequence with a top-3 accuracy of more than 70\% and a top-10 accuracy of around 90\%. Second, {\hmm} consistently outperforms $n$-gram model with a difference of around 10\%.

In summary, the core contributions of this paper include:

\noindent 1. {\model}, a graph-based representation of API usage scenarios;

\noindent 2. {\hmm}, a statistical, generative model of API usages; and
 
\noindent 3. Three algorithms to extract {\model} from apps' bytecode, to train {\hmm} based on method call sequences extracted from {\model}, and to recommend method calls in code completion engines using the trained {\hmm}.

\noindent 4. An empirical evaluation for the accuracy and usefulness of {\tool}

The remaining of this paper is organized as the following. Section II presents some examples of Android APIs and their usages in practice. Section III introduces our two main conceptual contributions: {\hmm} and {\model}. Section IV discusses the overall architecture of {\tool} and its main algorithms. Our empirical evaluation is presented in Section V. We discussed related work in Section VI and conclude this paper in the last section.

\section{Motivation}

\begin{figure}
	\centering
	\includegraphics[scale = 0.4]{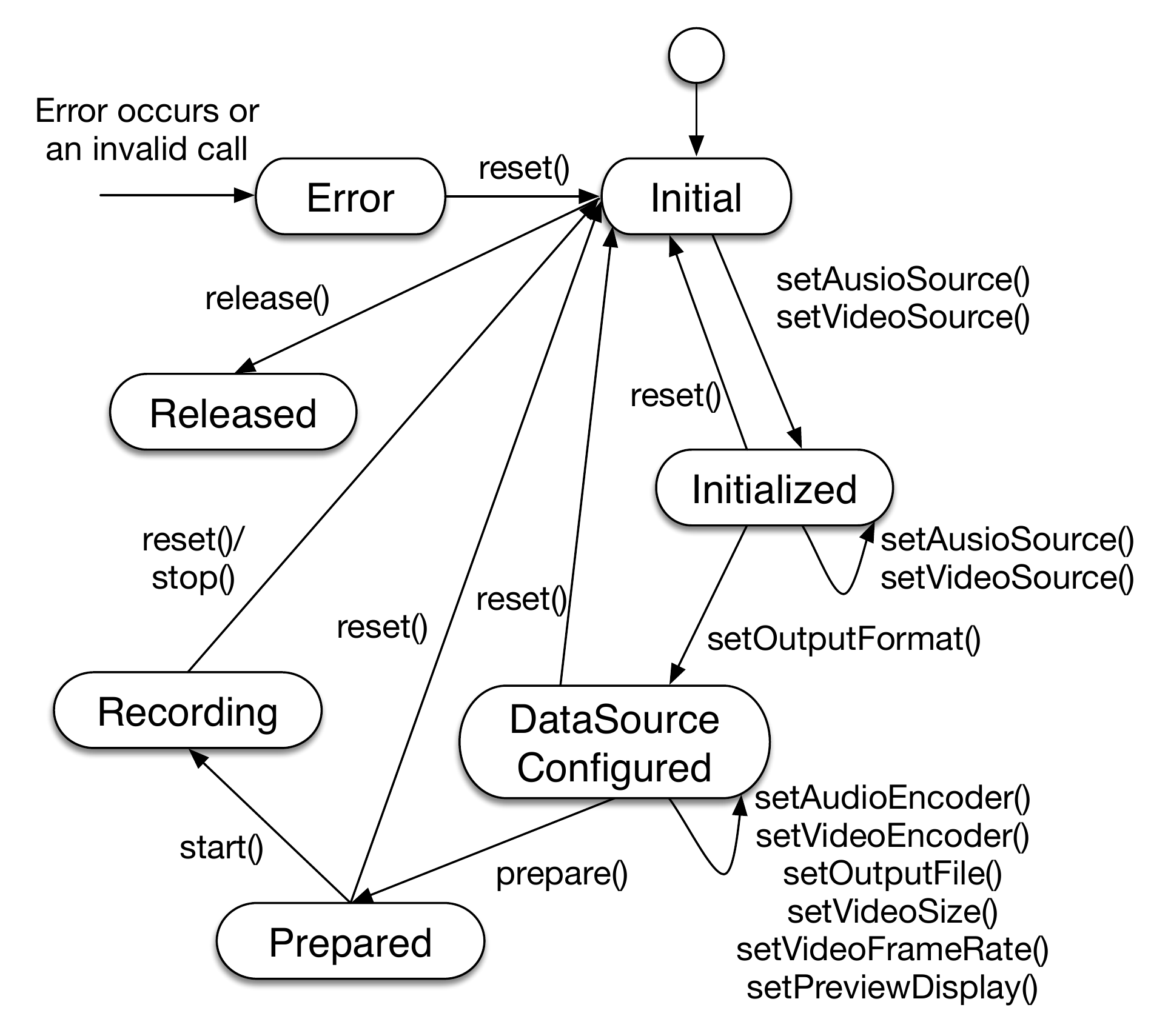}
	\vspace*{-0.4cm}
	\caption{The MediaRecorder State Diagram}
	\label{fig:mediarecorder_diagram}
	\vspace*{-0.5cm}
\end{figure}

\begin{figure}[t]
	\centering
\begin{lstlisting}[language=Java, basicstyle=\scriptsize\sffamily, frame=single]
protected void startRecording(String file) {
    MediaRecorder recorder = new MediaRecorder();
    recorder.setAudioSource(MediaRecorder.AudioSource.MIC);
    recorder.setOutputFormat(MediaRecorder.OutputFormat.THREE_GPP);
    recorder.setAudioEncoder(MediaRecorder.AudioEncoder.AMR_NB);
    recorder.setOutputFile(file);
    try {
        recorder.prepare();
    } catch (IllegalStateException e) {
        e.printStackTrace();
    } catch (IOException e) {
        e.printStackTrace();
    }
    recorder.start();
}
\end{lstlisting}
	\vspace*{-0.4cm}
	\caption{Source code examples of using \code{MediaRecorder} objects}
	\label{code:mediarecorder_examples}
	\vspace*{-0.8cm}
\end{figure}

In this section, we discuss an example to demonstrate the challenges of learning APIs in programming. Assume that a developer wants to program his app to record audio and video. He thinks of class \code{MediaRecorder} in the Android application framework but does not know how to write the code to initialize a \code{MediaRecorder} object and calls relevant methods. In other words, he has to learn the usages of this API object.

One way to learn such API usages is to read object state diagrams. Figure~\ref{fig:mediarecorder_diagram} illustrates the state diagram of \code{MediaRecorder}, which is reproduced from Android Developer website\footnote{http://developer.android.com/reference/android/media/MediaRecorder.html}. As seen in the figure, the state diagram of an object is a finite state machine in which each node represents a state of the object and each edge represents a method call. We learn from the state diagram that a \code{MediaRecorder} object has 7 states during its lifetime and several methods can be called to change from one state to another. For example, if the object is in state \code{Prepared} and method \code{start()} is called, it will change to state \code{Recording}. Although useful, state diagrams are often non-existed for most API objects. In addition, it is difficult to infer from the diagram the usages for a typical or popular task, such as the method sequences to set up a \code{MediaRecorder} for audio or video recording.

A more popular way to learn API usages is from reading real-world code examples. For example, Figure \ref{code:mediarecorder_examples} provides an example using a \code{MediaRecorder} object to record audio. Although more illustrative, this example also shows that this usage scenario of a \code{MediaRecorder} object for recording audio is rather complex. It involves 7 method calls, two exception handler, and several constants. More severely, for mobile API framework, code examples of most API usages are often unavailable or insufficient. Unlike in traditional software development where massive open source repositories are available, when developing mobile apps, developers often publish their apps to only online app stores. In other words, most mobile apps are closed-source, i.e. their source code is not available online for code search.

From this examples we see that programmers can learn API usages from state diagrams and code examples to use them correctly. However, those learning materials are often complex, unavailable, or insufficient, especially for mobile API framework. 


To address these challenges, in this paper, we propose HAPI (Hidden Markov Model of API usages), a statistical model for API usages that learns ``probabilistic'' state diagrams from a vast amount of publicly available bytecode. Our idea is based on an assumption that usage patterns specified by state diagrams exhibit on code examples. HAPI is adopted from Hidden Markov Models \cite{hmm_intro}, a probabilistic generative model for modeling sequences with ``hidden'' states. 

\section{Approach}

\begin{figure}
\centering
\includegraphics[scale = 0.4]{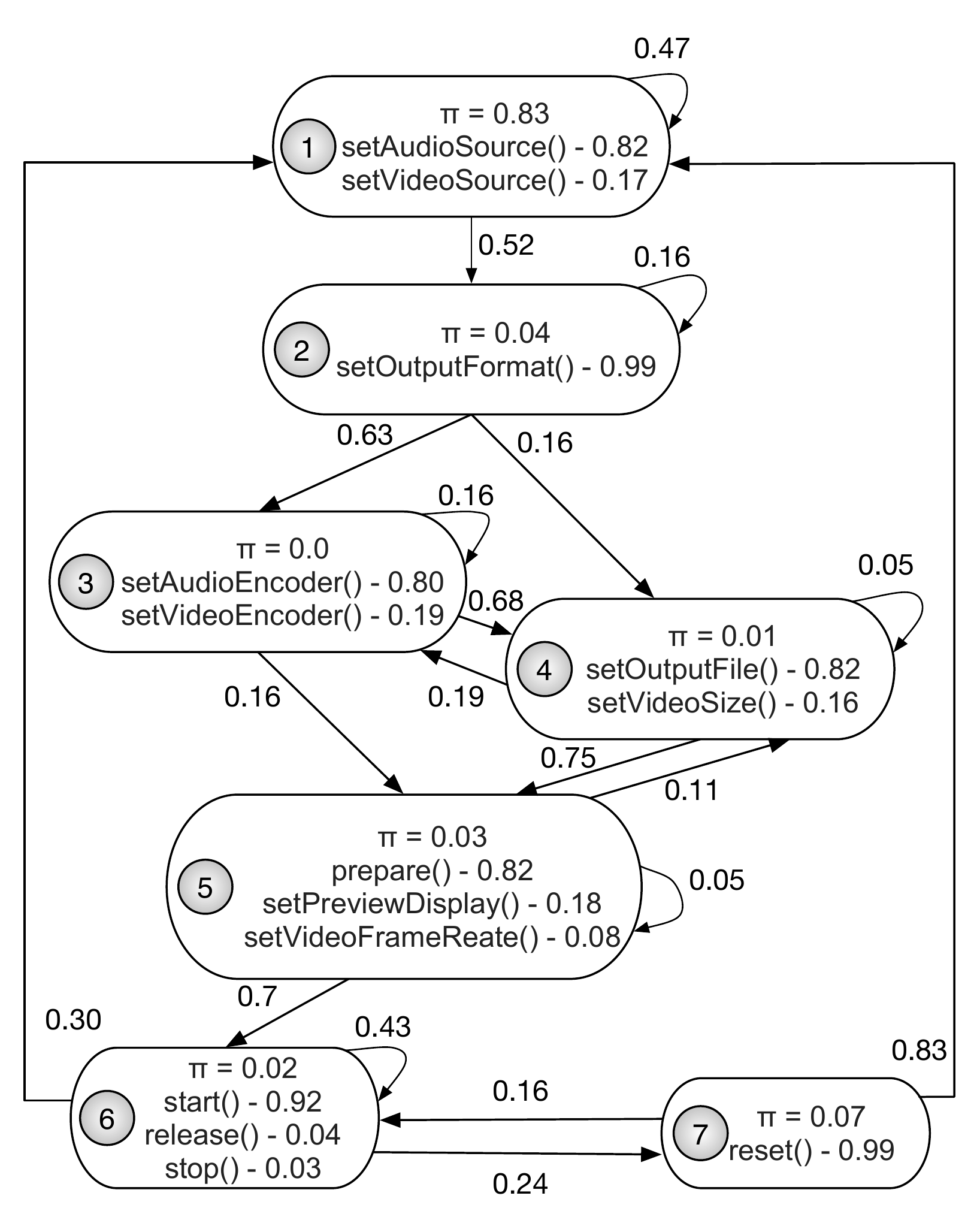}
\vspace*{-0.4cm}
\caption{The {\hmm} model to represent usage patterns of \code{MediaRecorder} object}
\label{mediarecorder_hmm}
\vspace*{-0.8cm}
\end{figure}

Figure \ref{mediarecorder_hmm} illustrates the ``probabilistic'' state diagram (HAPI) representing the usages of \code{MediaRecorder} object learned by our approach. As seen in the figure, each node in {\hmm} represents a state of the object as in the state diagram. But different from state diagrams, each state in {\hmm} has a probability to be selected as the starting state $\pi$ and a calling distribution which specifies the probability of calling a particular method at that state. An directed edge from state $x$ to state $y$ has a weight representing the probability of switching from state $x$ to state $y$ after calling a method. In the figure, to simplify the drawing, we only show edges with significant weights.

{\hmm} has several advantages over state diagrams. First, the state diagram only specifies the general rules for using objects but not specify the common usages. Once learned from code examples, {\hmm} could infer common usages of an object by searching for the method sequences that have highest probability. Second, {\hmm} could be used to recommend API usages when developers are programming. For example, given a sequence of API method calls, it can predict the most likely next method call. Let us describe it in details.

\subsection{Hidden Markov Models for learning API usages}

In this section, we present {\hmm}, a statistical model for API usages, which is adopted from Hidden Markov Models \cite{hmm_intro}. {\hmm} is a generative probabilistic model that describes the process of generating API method sequences of an API object (set of objects). To describe the model, we denote the API method invocation at time $t$ by the random variable $Y_t$. $Y_t$ is a variable follows the discrete distribution and $t$ is an integer-value index.

{\hmm} has two defining properties. First, {\hmm} has several internal states that represent states of the object (set of objects). Each state has a calling distribution which specifies the probability of calling a particular method on the state, and a transition distribution which specifies the probability of changing from that state to a particular state after calling a method. {\hmm} assumes that the method invocation on the object at time $t$ is sampled from the calling distribution of state $q_t$ which is hidden from observers. We could think that the hidden state $q_t$ is the current state of the object time $t$. Second, {\hmm} assumes that the state $q_t$ satisfies the \textit{Markov property}: that is, given the value of $q_{t-1}$, the probability of changing to $q_t$ is independent of all the states prior to $t-1$. In other words, the state at some time encapsulates all we need to know about the future of the generation process.

For a formal description, we will use the following notations to characterize a HAPI:
\begin{itemize}
\item $Q = \{q_1, q_2, ..., q_K\}$, the set of $K$ hidden states of an API object (set of objects)
\item $V = \{v_1, v_2, ..., v_M\}$,  the set of API methods associate with the object (set of objects).
\item $A = \{a_{ij}, a_{ij} = P(q_j \mbox{ at } t+1 | q_i  \mbox{ at } t)\}$, the state transition matrix, where the state transition coefficients satisfy $\sum_{j} a_{ij} = 1$, $a_{ij}$ is transition probability from state $q_i$ to state $q_j$
\item $B = \{b_{j}(m)\}, b_{j}(m) = P(v_m \mbox{ at } t | q_j  \mbox{ at } t)$, $b_{j}$ is the calling distribution in state $q_j$, $b_{j}(m)$ is the probability of calling method $v_m$ in the state $q_j$.
\item $\pi = \{ \pi_{i} \}, \pi_{i} = P(q_i \mbox{ at } t  = 1)$, the initial state distribution, where $\pi_{i}$ is the probability of that state $q_i$ is the beginning state to generate a method sequence.
\end{itemize}

Given the {\hmm} model of an object (a set of objects) with parameters $\lambda = (A, B, \pi)$, the generative process of an method sequence $Y = (y_1, y_2, .., y_T)$ is as follows:\\
1. Sample an initial state $i_1$ from the initial state distribution $\pi$ and set $t = 1$\\
2. Sample an method call $y_t$ from the calling distribution of $i_t$,  $b_{i_t}(m)$\\
3. Sample $i_{t+1}$ from the transition distribution of  $i_t$, $a_{i_t i_{t+1}}$\\
4. Set $t = t + 1$ and return to step 3 if $t < T$; otherwise end.

Using {\hmm} to model API usages, in our work, we are interested in 1) how to learn {\hmm} model of an API object (set of objects), 2) how to make used of learned {\hmm} models. For the first problem, we present an algorithm to train a {\hmm} (estimating its parameter $\lambda = (A, B, \pi)$) given collection of method sequences involvings the API object (set of objects). For the second problem, in our work, we focus on using {\hmm} for API usage recommendation. We design an algorithm to use the trained {\hmm} to compute the generating probabilities of several method sequences to rank those sequences and use the ranking for recommendation task.

\subsection{Abstract Representation of Usage Scenario}


Input to train {\hmm} model is API method sequences associate with single object or a set of objects involves in an usage. To extract such sequences, we propose a novel graph-based representation called Abstract Representation of Usage Scenario ({\model}). {\model} is a labeled, directed graph representing an usage scenario in one execution path including many objects, methods, complex control, and data flows between them.

\begin{definition}[{\model}]
{\model} is a directed graph that has two types of nodes and edges. Object nodes represent objects that are involved or created during the execution. Action nodes are any actions that performs on object nodes. Action nodes could be object instantiations, method calls, data field accesses, or other operations. Control edges represent temporal orders between action nodes. Data edges describe data-dependency between object and action nodes.
\end{definition}

In {\model}, each object created or involved during the execution is represented as an object node. We also treat primitive variables as object nodes. Action nodes represent any action that is performed on object nodes. Action nodes could be object instantiations, method calls, data field accesses of one object, or other operations. Object nodes are labeled by class names (object nodes represent primitive variables are labeled by types). Action nodes of types object instantiations, method calls, or data field accesses are labeled as ``C.m'' with C is its class name and m is the method (or field) name. Other action nodes that represent operations are labeled as the name of the operation.

The control edges of {\model} are used to represent the temporal orders between action nodes. A control edge from an action node A to action node B means that A is executed before B in the execution path. Because {\model} is defined for each execution path, thus, there is only one temporal order between action nodes, which is represented by a set of control edges between action nodes. The data edges indicate the data dependencies between data nodes and action nodes. A data edge from object node A to action node B means that A is a parameter of the action that B represents. A data edge from an action node B to data node A means the action B returns object node A as a result.

Figure \ref{DOUMExample1} shows two {\model}s of a method. Rectangle nodes are action nodes, while object nodes are represented as round rectangle nodes. Solid arrows represents control edge between action nodes and dashed arrows represent data edges.

\begin{definition}[Usage-dependency]
A set of object nodes is said to have usage-dependency if there is at least one action node that have data-dependent with all the object nodes in the set.
\end{definition}

There are two main advantages of representing usage scenario using {\model}. First, we could easily extract action sequences associates with one or a set of objects. For a object node (set of object nodes), we identify all action nodes that have data-dependency with that object node (set of object nodes) based on data edges. The action sequence follows that temporal order between action nodes. In case we want to extract API method sequences we filter out all non API action nodes in action sequences. The second advantage of {\model} is that we could define whether multiple object nodes join in an usage or they are used in several usages separately. We define a set of object nodes usage-dependency if there is at least one action node that have data-dependency with all the object nodes in the set. By defining usage-dependency we could model API usages that involves several objects.

\section{System Description}
\begin{figure*}
\centering
\includegraphics[scale = 0.45]{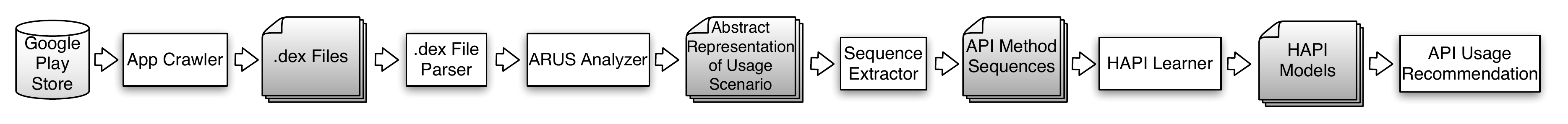}
\vspace*{-0.2cm}
\caption{The overview of {\tool}}
\label{overview}
\vspace*{-0.8cm}
\end{figure*}

Our approach for learning API usages consists six major components: app crawler, .dex file parser, {\model} analyzer, sequence extractor, {\hmm} learner,  and API usage recommendation. Figure \ref{overview} shows an overview of all the components in the system. Our approach consists of three main phases. In app processing phase, the app crawler downloads (free) Android applications from the Google Play Store and extract the .dex files (Dalvik Execution Format) that contain bytecode of those apps to store in local repository. The .dex file parser reads .dex files and construct Control Flow Graph (CFG) for bytecode of each method. In analyzing phase, {\model} analyzer takes a CFG as an input and produces {\model} for each execution path in the CFG. Sequence extractor uses these produced {\model}s and extracts API method sequences of one or multiple objects. In the learning phase, the  {\hmm} learner uses the API method sequences that have been extracted to learn API usage models. For evaluation, we used our {\hmm} models for API usage recommendation. Next, we present the details for each component in the system. 

\subsection{App Crawler}
In {\tool}, we first built a crawler to downloads applications from Google Play Store. We developed our app crawler based on the google-play-crawler project\cite{googleplay_crawler}. The crawler simulates an android devices and it has the same accessibility to the store as a normal phone such as searching for apps, browsing categories and apps, app information, downloading free apps, etc. On Google Play Store, each category contains top lists of apps. We downloaded all the top free/new free apps in all categories. In each application package (.apk files), we only kept the .dex files that contain all the bytecode of the applications. 

\subsection{.dex File Parser}

The next step of data processing phase is parsing .dex files to obtain bytecode instructions of methods. For this task we use smali tool \cite{smali}, which is an assembler/disassembler for Android's dex format. After obtaining the bytecode instructions for methods, to make it easier to analyze, we contruct Control Flow Graph (CFG) from those instructions. 

\noindent\textbf{Dalvik Virtual Machine.} Android applications are developed in Java, but unlike a normal Java application, which is compiled into Java Virtual Machine (JVM) class files and executed in JVM, Android applications are compiled to Dalvik bytecode and are executed in Dalvik Virtual Machine (DVM). DVM is a virtual machine that provides runtime environment for application in Android platform. Dalvik bytecode of all methods implemented in an Android application is stored in a single .dex file inside the application package.

DVM is a register-based virtual machine. DVM uses a set of 32-bit registers to hold bit values (such as integers and floating point numbers) and object references. Frames (activation records) of  a method are fixed in size which consists of a particular number of register. These registers are used to store local variables, parameters, return values, and temporary values instead of using stacks as in Java Virtual Machine. The $N$ arguments to a method are stored in the last $N$ registers of the method's frame. Instance methods are passed a \code{this} reference as their first argument. 

Figure \ref{bytecode_example} shows Dalvik bytecode for the \code{readTextFile} method. The method is allocated 7 registers \code{v0-v6} to executes instructions. Registers \code{5} and \code{v6} store parameters, \code{v5} is the object that the method is being invoked (because \code{startRecording} is an instance method), \code{v6} is a string contains file name. The registers \code{v0-v4} are used to store local or temporary variables. Instructions in Dalvik bytecode operate on registers. For example, \code{mul-int v2,v5,v3} instruction multiples the values of registers \code{v2} and \code{v5} and stores the result to \code{v3}, or \code{new−instance v1, Ljava/io/FileReader;} creates a new \code{FileReader} object and returns the reference of that object to register \code{v1}.


\noindent\textbf{Control Flow Graph.} We construct Control Flow Graph (CFG), which is a representation of bytecode instructions. 
A node in the constructed CFG contains a single bytecode instruction and an edge is the control flow between the two instructions.
There are two types of nodes in CFG. Control nodes represent control instructions in bytecode, e.g. \code{if}, \code{return}, \code{throw}, or \code{goto} instructions. Other instructions are normal nodes. Normal nodes in a CFG only have one out-going edge points to the next instruction while control nodes could have several out-going edges (\code{if}, \code{switch} nodes) or do not have any (\code{return} nodes). Techniques used for constructing CFGs are quite standard. First, we create all nodes in the CFG, each one corresponds to an instruction in the instruction list. Then, for each node, we use offsets to identify nodes that are executed after it and add edges from the current node to those nodes.

\begin{figure}[h]
	\centering
\begin{lstlisting}[language=Java, basicstyle=\scriptsize\sffamily, linewidth = \linewidth, frame = single]
public String readTextFile(String fileName) throws IOException {
	FileReader fr = new FileReader(fileName); 
	BufferedReader br = new BufferedReader(fr);
	StringBuilder strBuilder = new StringBuilder(); 
	String line;
	while((line = br.readLine()) != null) { 
		strBuilder.append(line);
	}
	br.close();
	return strBuilder.toString();
}
\end{lstlisting}
	\vspace*{-0.2cm}
	\caption{A source code example}
	\label{source_example}
	\vspace*{-0.8cm}
\end{figure}
\begin{figure}[h]
	\centering
\begin{lstlisting}[language=Java, basicstyle=\tiny\sffamily, frame = single, linewidth = \linewidth]
|[0001dc] IOManager.readTextFile:(Ljava/lang/String;)Ljava/lang/String;
|0000: new-instance v1, Ljava/io/FileReader;
|0002: invoke-direct {v1, v6}, Ljava/io/FileReader;.<init>:(Ljava/lang/String;)V
|0005: new-instance v0, Ljava/io/BufferedReader;
|0007: invoke-direct {v0, v1}, Ljava/io/BufferedReader;.<init>:(Ljava/io/Reader;)V
|000a: new-instance v3, Ljava/lang/StringBuilder;
|000c: invoke-direct {v3}, Ljava/lang/StringBuilder;.<init>:()V
|000f: invoke-virtual {v0}, Ljava/io/BufferedReader;.readLine:()Ljava/lang/String;
|0012: move-result-object v2
|0013: if-nez v2, 001d
|0015: invoke-virtual {v0}, Ljava/io/BufferedReader;.close:()V
|0018: invoke-virtual {v3}, Ljava/lang/StringBuilder;.toString:()Ljava/lang/String;
|001b: move-result-object v4
|001c: return-object v4
|001d: invoke-virtual {v3, v2}, Ljava/lang/StringBuilder;.append:(Ljava/lang/String;)Ljava/lang/StringBuilder;
|0020: goto 000f
\end{lstlisting}
	\vspace*{-0.2cm}
	\caption{Dalvik bytecode for source code example in Figure \ref{source_example}}
	\label{bytecode_example}
	\vspace*{-0.8cm}
\end{figure}

\subsection{{\model} Analyzer}
{\model} analyzer takes CFG as input and produce {\model}s which describes usage scenario for each execution path. 
\begin{figure}[h]
\centering
\begin{lstlisting}[numbers=left,basicstyle=\scriptsize\sffamily, frame=single,xleftmargin=2.5em,framexleftmargin=2.5em]
function BuildARUS(Method $M$)
  $CFG$ = BuildCFG($M$)
  $A = \emptyset$  //list of ARUSs
  $S_{t}$ = CreateStartState($CFG, M$)
  $F = \emptyset$
  Push($F, S_{t}$)
  while $F \neq \emptyset$ 
    $S$ = Pop($F$)
    $SN$ = GetStartNode($S$)
    while $SN$ is not a control node 
      BuildTemporaryARUS($S, SN$)
      AddExploredNode($S, SN$)
      $CN$ = GetNextNode($SN$)
    if $SN$ is the return node
      AddARUS($A, S$)
    else 
      for $NN \in$ GetNextNodes($SN$)
        if $NN \notin$ GetExploredNodes($S$)
          $S_{c}$ = GetCopy($S$)
          SetStartNode($S_{c}, NN$)
          Push($F, S_{c}$)
  return $A$
\end{lstlisting}
\vspace*{-0.4cm}
\caption{Building {\model} Algorithm}
\label{aus_algorithm}
\vspace*{-0.6cm}
\end{figure}

\subsubsection{The Algorithm}
The main idea of the algorithm to construct {\model} is to explore all the execution path in CFG and build temporary {\model}s when exploring those paths. Once a path has been explored, it collects the {\model} that have been built for that path. Each CFG has a start node which are the first instruction and a termination node which is the return node. Our algorithm needs to find all execution paths from the start node to the termination node. One problem that the algorithm needs to consider is to handle loops that occur in the CFG. These loops represents \code{while} or \code{for} loops in source code. The instructions inside loops may be executed either several times or zero time, thus, could lead to infinity number of execution paths. On the other hand, considering these instructions once can help build usage scenario associated with loops. Therefore, our algorithm treats instructions inside a loop are only executed either once or not. In other words, a loop is considered to executed at most one time. 

\textbf{Detailed Algorithm.} Figure \ref{aus_algorithm} show the pseudo-code for our algorithm. Input of the algorithm is the bytecode of a method $M$. The algorithm starts with creating $CFG$ from the bytecode instructions using techniques described the previous section (line 1).  Similar to depth first search algorithm, we maintain a stack $F$ to store states which are frontiers to explore (line 4). Each state of our algorithm represents an incomplete execution path and contains following information: 1) start node: the current node in $CFG$ when a state is pop from stack $F$, 2) explored nodes: a set of all nodes in CFG that have been visited 3) a temporary {\model}. The initial state $S_t$ is created in line 3 with start node is the begin node of $CFG$, explored nodes and the temporary {\model} of this state are empty. The stack $F$ is initialized with $S_t$ in line 4. In the main \code{while} loop of the algorithm, each time, a state $S$ is pop from $F$. We start with $SN$, which is the start node of $S$ (line 8) and explore the path from $SN$ to the next control node in $CFG$. Whenever $SN$ is a normal node in $CFG$, we use $SN$ to build and update the temporary {\model} of $S$ (we will describe the algorithm to build a temporary {\model} in the next section). We then add $SN$ to the set of explored nodes of $S$ and update $SN$ equal to the next node of $SN$ in $CFG$ (because $SN$ is still a normal node in $CFG$, it only has one next node). After the loop from line 9 to line 12. $SN$ now is a control node in $CFG$. If $SN$ is the return node of $CFG$ then we have explored one execution path of $CFG$, the temporary {\model} of $S$ now becomes a final {\model}, thus, it is added to the list $A$ (line 14). If $SN$ is not the return node, we need to consider all unexplored out going edges of $SN$. For each next node $NN$ of $SN$, if $NN$ is not in the set of all explored nodes of $S$, we copy all information of $S$ into a new state $S_c$, set $NN$ is the start node of $S_c$, and push $S_c$ into $F$. The algorithm terminate when the stack $F$ becomes empty. Our algorithm explored all the paths of $CFG$ and only go through each loop at most one time. The list $A$ now contains all {\model} of the method $M$. We next describe algorithm to build temporary {\model}.

\subsubsection{Building temporary {\model}}

\begin{figure}
	\centering
	\includegraphics[scale = 0.4]{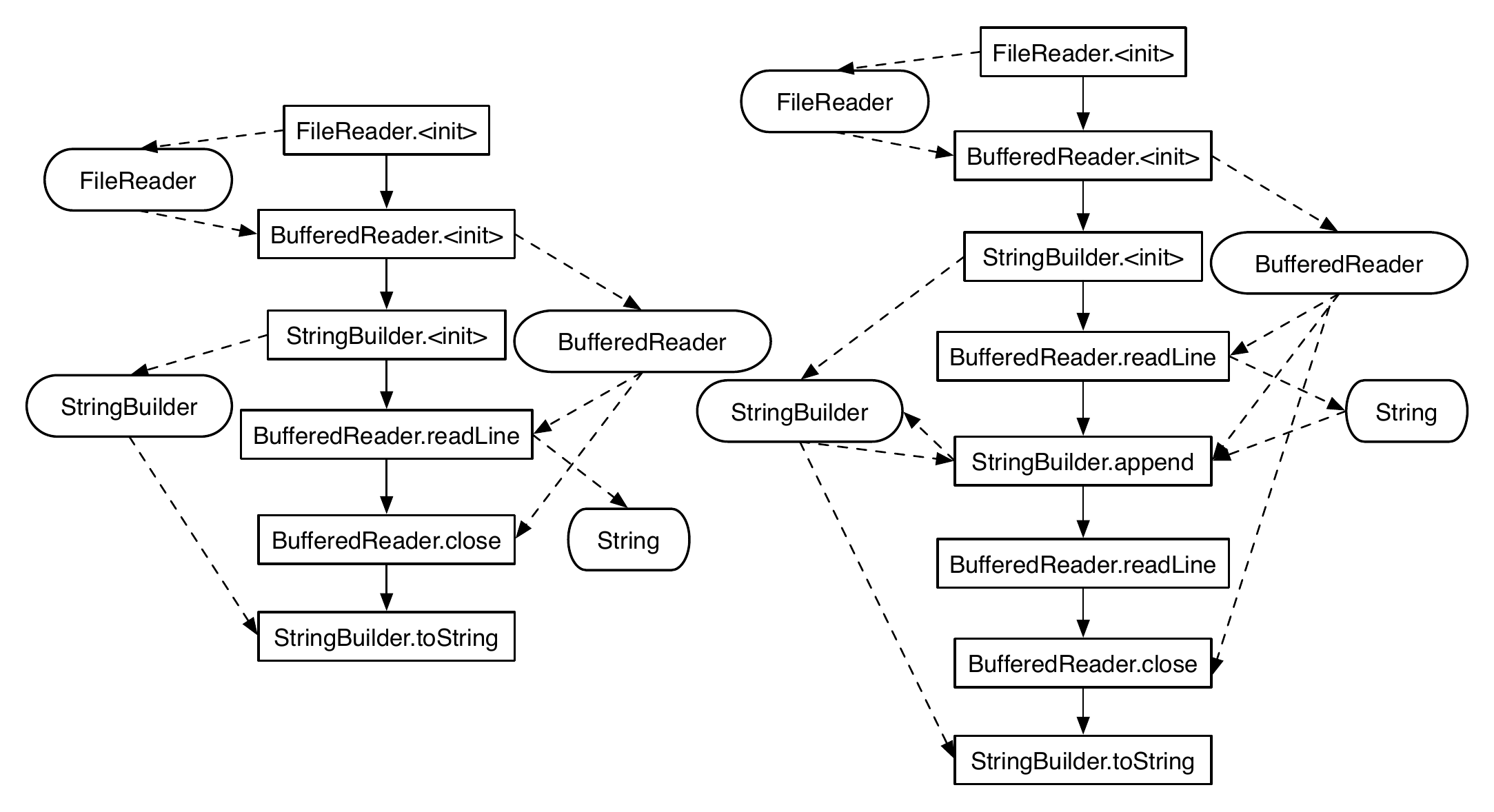}
	\caption{{\model}s}
	\label{DOUMExample1}
	\vspace*{-0.8cm}
\end{figure}

In this section, we describe our technique to build and update the temporary {\model}. Initially, before considering any instructions, we create an empty {\model}. For each parameter of the method we create a corresponding object node and add to the {\model}. When an instruction is considered (line 14), it create a new action node and add that node to the current {\model}. After adding the new action node, we need to update edges. We add a control edge from the last inserted action node to the new action node. For data edges, we first need to add data edges from object nodes that are parameters of the instruction to the new action node. To do that, we maintain a map $M_{current}$ to store the current mapping between registers and object nodes. When an action node is created, we use $M_{current}$ to identify all object nodes that are used as parameters of the action, and we add data edges from those object nodes to the action node. Consider the  instruction \code{000f: invoke-virtual {v0}, Ljava/io/BufferedReader;.readLine:() Ljava/lang/String;}, it has an input parameter, which is the register \code{v0}, using $M_{current}$ we know that when we create an action node from the instruction, v0 is currenty holding a reference to a BufferedReader object node. We then add a data edge from the object node to the newly created action node. An instruction may return a value to a register. If it does not create a new object i.e. it returns a reference of an object has already been created before, then we identify the object node represent that object by $M_{current}$ and add a data edges from the action node to this node. Otherwise, we create a new object, add a data edge from the action node to the new object node and update mappings in $M_{current}$.

\subsection{Sequence Extractor}

%
In this section, we describe how to extract API method sequences for one object or multiple objects from of a method using {\model}.
For single object, sequence extractor scans through an {\model} to find object nodes associates with Android API objects. 
For each object node $O_i$ we find all action nodes that have data edges connected with $O_i$ and sort these action nodes by execution order. 
From the sorted action sequence, we only consider action nodes that represent API methods to get an API method sequence.
For example, in Figure \ref{DOUMExample1}, for the first {\model}, there are three action nodes related to the object node \code{BufferedReader} and all of them are Android API methods, thus, the API method sequence associates the object node \code{BufferedReader} is \code{(BufferedReader.init, BufferedReader.readLine, BufferedReader.close)}. 

To extract action sequences that involve multiple API objects, we first identity usage-dependent object sets. For each action node that represents API methods, we collect all API object nodes that have data-dependent with it, then we form the set of object types corresponds to those object nodes. In the first {\model} of Figure \ref{DOUMExample1}, there are two sets of usage-dependent objects: \code{(FileReader, BufferedReader)} has data-dependency with the action node \code{BufferedReader.init}, and \code{(BuffedReader, String)} has data-dependency with the action node \code{BufferedReader.readLine}.After collecting usage-dependent object sets, for each object set, we extract corresponding API method sequences using same technique for single object. For example, the API method sequence for the set \code{(FileReader, BufferedReader)} in the first {\model} is \code{(FileReader.init, BufferedReader.init, BufferedReader.readLine, BufferedReader.close)}. 

There is one method sequence for an object (set of usage-dependent objects) in each {\model} of a method. Thus, after extracting method sequences in all {\model}s of the method, we remove duplicate sequences and only keep distinct sequences for each object (set of usage-dependent objects).


\subsection{{\hmm} Learner}
The {\hmm} learner uses a collection of API method sequences of an object (a set of usage-dependent objects) to learn the API usages model. In this section, we describe the training algorithm to estimate parameters of {\hmm} model from training data. We also present a method to choosing the number of hidden states.
\subsubsection{Training Algorithm}
The training algorithm aims to estimate {\hmm}'s parameters $\lambda = (A, B, \pi)$  given a collection of API method sequences. 
In general, Baum-Welch algorithm is often used to estimate parameters of Hidden Markov Model \cite{hmm_intro}. In this paper, we present a modified version of Baum-Welch algorithm for the problem of learning API usages. The input of the algorithm is a collection of API method sequences. We observed that there are many method sequences are duplicated in the colllection. Thus, to save space and speed up the training algorithm, we store training data as a map, where each method sequence is mapped to the number of time it occurs in the collection. Initially, the parameters of a {\hmm} are assigned with random values. The main idea of the algorithm is to iteratively estimates parameters to maximize the likelihood function (the probability of generating data given model). The iterative process terminates when the estimated values of parameters converge.

\begin{figure}[t]
\begin{lstlisting}[basicstyle=\scriptsize\sffamily, mathescape, numbers=left,frame=single,xleftmargin=2.5em,framexleftmargin=2.5em, deletekeywords={forward}]
function TrainHAPI(TrainSet $S$, NHiddenStates $K$)
  initialize $\lambda = (A, B, \pi)$ with random values
  repeat
    foreach $( Y_n = (y_1, y_2, ..., y_T), c_n) \in S$
      $\alpha$ = Forward($\lambda, Y_n, T$)
      $\beta$ = Backward($\lambda, Y_n, 1$)
      for $i \in 1..K$: // Compute state and state transition probabilities
        $\gamma_{i}^{n}(t) =  \frac{\alpha_{i, t}\beta_{i, t}}{\sum_{j=1}^{K}\alpha_{j, t}\beta_{j,t}}$
        for $j \in 1..K, t \in 1..T-1$:
          $\xi_{ij}^{n}(t) = \frac{\alpha_{i, t}a_{ij}\beta_{j, t+1}b_j(y_{t+1})}{\sum_{k=1}^{K}\alpha_{k}(t)\beta_{k, t}}$
    for $i \in 1..K$:  //update model parameters
      $\pi_{i}^{(s+1)} = \frac{1}{D} \sum_{n=1}^{N} c_n \times \gamma_{i}^{n}(1)$
      for $j \in 1..K$:
        $a_{ij}^{(s+1)} = \frac{\sum_{n=1}^{N} c_n \times \sum_{t=1}^{T-1} \xi_{ij}^{n}(t)}{\sum_{n=1}^{N} c_n \times \sum_{t=1}^{T-1} \gamma_{i}^{n}(t)}$
      for $v_m \in V$:
        $b_{i}^{(s+1)}(v_m)  = \frac{\sum_{n=1}^{N} c_n \times  \sum_{t=1}^{T} 1_{y_t = v_m} \xi_{i}^{n}(t)}{\sum_{n=1}^{N} c_n \times \sum_{t=1}^{T} \gamma_{i}^{n}(t)}$
  until convergence
  return $\lambda$

function Forward(HAPI $\lambda$, APISequence Y, Position P)
  $\alpha = \{\alpha_{i,t}\}\quad 1 \leq i \leq K, 1 \leq t \leq P$
  for $i \in 1..K$: $\alpha_{i,1} = \pi_{i}b_{i}(y_{1})$
  for $i \in 1..K, t \in 2..P$:
      $\alpha_{i,t} = b_i(y_{t})\sum_{j=1}^{K}\alpha_{i,t-1} a_{ij}$
  return $\alpha$

function Backward(HAPI $\lambda$, APISequence Y, Position P)
  $\beta = \{\beta_{i,t}\} \quad 1 \leq i \leq K, P \leq t \leq T$
  for $i \in 1..K$: $\beta_{i,T} = 1$
  for $i \in 1..K, t \in T-1..P$:
      $\beta_{i,t} = \sum_{j=1}^{K}\beta_{j,t+1}a_{ij}b_{j}(y_{t+1})$
  return $\beta$
\end{lstlisting}
\vspace*{-0.4cm}
\caption{Training Algorithm}
\label{trainhapi}
\vspace*{-0.8cm}
\end{figure}

Figure \ref{trainhapi} shows the algorithm for training {\hmm}. its input includes training data $S$ and the number of hidden states $K$. The parameters of the model are initialized randomly (line 1). Let us describe steps in the main loop of the algorithm:

\noindent\textbf{Step 1}. Compute forward and backward probabilities. For each method sequence in training data we use dynamic programming to compute forward probabilities $\alpha_{i,t}$ (using \code{Forward} function) and backward probabilities $\beta_{i,t}$ (using \code{Backward} function)~\cite{hmm_intro}. $\alpha_{i,t}$ is defined as the probability of seeing partial method sequence $y_1,y_2,...,y_t$ and being in state $q_i$ at time $t$ given the model $\lambda$:
\small
\begin{equation}
\alpha_{i,t} = P(y_1, y_2, ..., y_t, i_t = q_t | \lambda)
\end{equation} 
\normalsize
$\beta_i(t)$ is the probability of seeing the ending partial sequence  $y_{t+1},...,y_{T}$ given state $q_i$ at time $t$ and the model $\lambda$:
\small
\begin{equation}
	\beta_{i, t} = P( y_{t+1},...,y_{T} | i_t = q_i, \lambda)
\end{equation}
\normalsize

\noindent\textbf{Step 2}. Compute state probabilities $\gamma$ and state transition probabilities $\xi$ using forward and backward probabilities. $\gamma_{i}^{n}(t)$ is probability of being at state $q_i$ at time $t$ given method sequence $Y_n$ and the model $\lambda$:
\small
\begin{equation}
	\gamma_{i}^{n}(t)  = P(i_t = q_t| Y_n, \lambda) = \frac{\alpha_{t,i}\beta_{i, t}}{P(Y|\lambda)}
\end{equation}
\normalsize
The state transition probability $\xi_{ij}^{n}(t)$ is the probability of being in state $q_i$ at time $t$ and making transition to state $q_j$ at time $t+1$ given method sequence $Y$ and the model $\lambda$: 
\small
\begin{equation}
	\xi_{ij}^{n}(t) = P(i_t = q_i, i_{t+1} = q_j | Y_n, \lambda)
\end{equation}
\normalsize
\noindent\textbf{Step 3}. Reestimate model parameters. In this step, the model parameters are estimated as expected values of  probabilities that we computed in the previous step.  $\gamma_{i}^{n}(t)$ is computed as the expected number of time state $q_i$ is at time 1. $a_{ij}^{(s+1)}$ is estimate as the expected number of transitions from state $q_i$ to state $q_ j$ compared to the expected total number of transitions from state $q_i$.  $b_{i}^{(s+1)}(v_m)$ is the expected number of times the output method have been equal to $v_m$ while in state $i$ over the expected total number of times in state $q_i$. In the update equations, $D$ is the total number of method sequences in the training set, and $1_{y_t = v_m}$ is the indicator function.

\subsubsection{Choosing the number of hidden states}

\begin{figure}
	\centering
	\includegraphics[scale = 0.3]{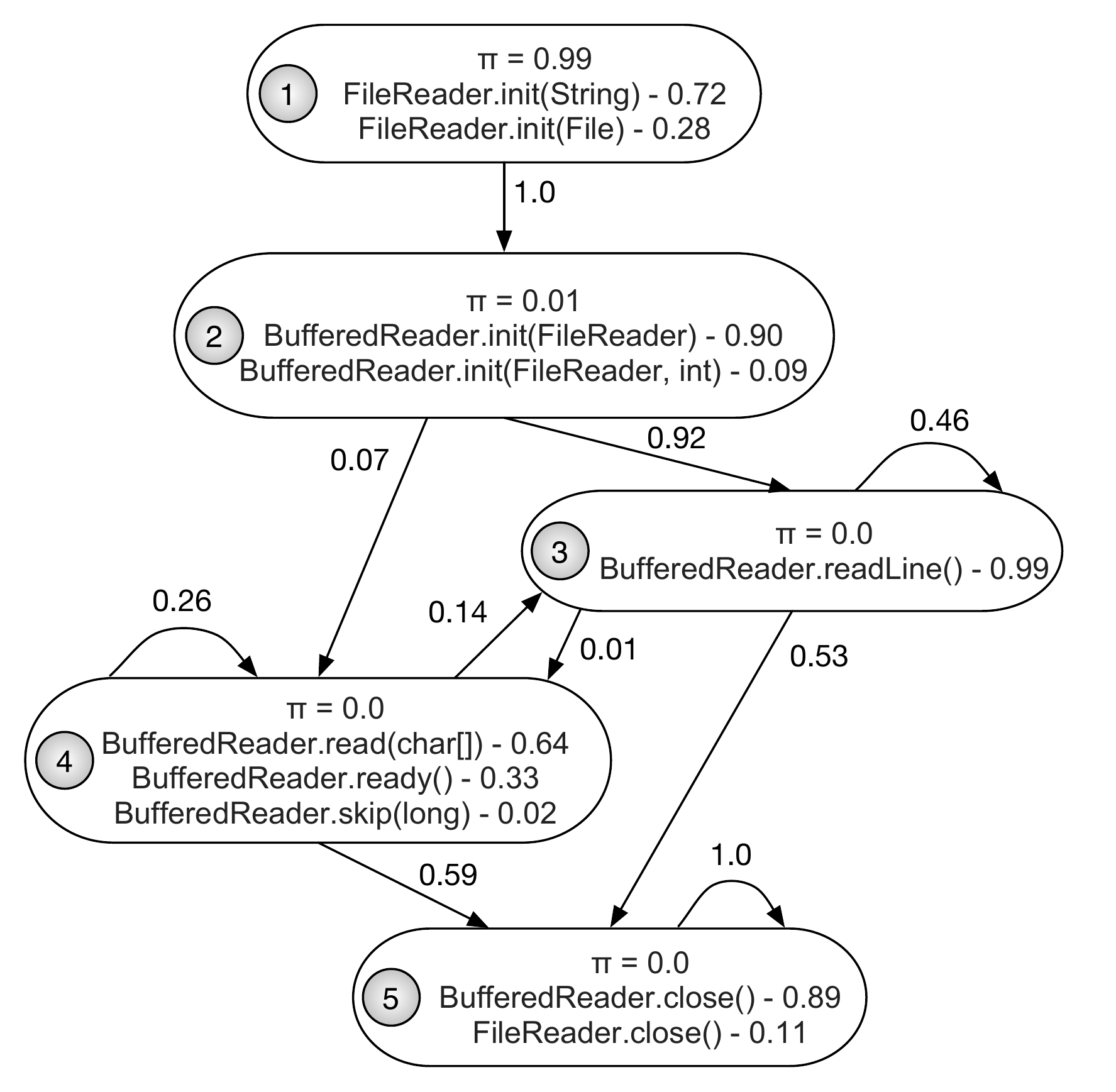}
	\vspace*{-0.3cm}
	\caption{{\hmm} for pair of objects \code{FileReader} and \code{BufferedReader}}
	\label{hapi_example}
	\vspace*{-0.5cm}
\end{figure}

When training the {\hmm} model for an object (or a set of objects), we need to specify number of the hidden states as an input for the training algorithm, but we often do not know the how many states that the object has.  Our idea in choosing the number of hidden states $K$ is try to build models with different $K$ and find a model that best describe new method sequences. This problem is equivalent to find $K$ that maximizes the probability of generating new data, i.e. likelihood function.
In this method, we divide a training set in to two sets, one is used to train models and one is used as validation data to optimize the number of hidden states. 
For each $K$ in a range, we train a HMM model with $K$ as the number of hidden states.
Then, for each model, we compute the likelihood of validation data which is the probability of generating the validation data given the model.
We then choose $K$ of the model that maximize the likelihood of the validation set.
Figure \ref{hapi_example} shows graph representation the {\hmm} model for the pair of objects \code{FileReader} and \code{BufferedReader} as a result of {\hmm} learner component. To make to display, we round probabilities and ignore probabilities that is less than 0.01.

\subsection{API Usage Recommendation}

In this section, we present an algorithm for recommending API method call using {\hmm}. The input of the algorithm is the {\hmm} model $H$ of an object (set of objects) and the incomplete API method calls $Y = (y_1, ..., y_N)$ associated with the object (set of objects) in current editing code. The location $T$ of $Y$ is missing. The output of the suggestion algorithm is a ranked list of API methods that could be filled as the method call at position $T$. The idea of our algorithm is to place each API method as the method call of $Y$ at location $T$ and compute score of this assignment as the probability of generating the updated sequence (including the new API method) given the {\hmm} model. Then we add the API method with score to the ranked list of all candidates. 

\begin{figure}[h]
	\begin{lstlisting}[basicstyle=\scriptsize\sffamily, mathescape, numbers=left,frame=single,xleftmargin=2.5em,framexleftmargin=2.5em, deletekeywords={forward}]
	function NextAPICall(HAPI $\lambda$, APISequence $Y$, Location $T$) 
	$R = \emptyset$ // a ranked list of candidates
	$\alpha$ = Forward($\lambda, Y, T-1$)
	$\beta$ = Backward($\lambda, Y, T+1$)
	for $v \in V$:
	for $i \in 1..K$:
	$\alpha_{i,T} = b_i(v)\sum_{j=1}^{K}\alpha_{j,T-1} a_{ij}$
	$\beta_{i,T} = \sum_{j=1}^{K}\beta_{j,T+1}a_{ij}b_{j}(v)$
	$score = \sum_{i=1}^{K} \alpha_{i,T} \beta_{i,T}$ 
	UpdateCandidateList($R, v, score$)
	return $R$
	\end{lstlisting}
	\vspace*{-0.4cm}
	\caption{Algorithm for suggesting the next API method call}
	\label{algo:hidden_states}
	\vspace*{-0.6cm}
\end{figure}

Figure \ref{algo:hidden_states} shows the algorithm. In the first part of our algorithm, we compute forward probabilities at position $T-1$ (using \code{Forward} function). We also compute backward probabilities at position $T+1$ (using \code{Backward} function). Then, we place each API method $v$  as the method call of $Y$ at position $T$ and compute forward and backward probabilities at that position (line 6-8). The $score$ of $v$ is the probability of generating sequence $(y_1, ..., y_T = v, ..., y_N)$ is computed by summing all product of forward and backward probabilities:
\small
\begin{equation}
P(Y, y_T = v|\lambda) = \sum_{i = 1} ^{K} P(Y, y_T = v, i_{T} = q_i | \lambda) =  \sum_{i=1}^{K} \alpha_{i,T} \beta_{i,T} \nonumber
\end{equation}
\normalsize
The algorithm returns a ranked list of all the API method candidates with scores for suggestion.

\section{Empirical Evaluation}
In this section we conducted several experiments to evaluate our approach. The first set of experiments aims to study the run-time performance of {\tool} in extracting method sequences from bytecode and in training its API usage models. The second set aims to compare the effectiveness of those usage models to the baseline model $n$-gram in recommending API method calls. All experiments are executed on a computer running 64-bit Ubuntu 14.04 with Intel Core i5 CPU 3.2Ghz processor, 4GB RAM, and a 500 GB drive storage.
\subsection{Collecting Data}
\begin{table}[t]
\caption{Data Collection}
\centering
\label{table:data_collection}
\begin{tabular}{lr}
\toprule
Number of apps analyzed                     & 22,330 \\
Number of classes analyzed                         & 4,861,196 \\
Number of methods analyzed                         & 9,117,943 \\
Number of bytecode instructions analyzed & 258,726,894 \\
Space for storing apps' bytecode (.dex files)				  & 68 GB\\
\bottomrule							
\end{tabular}
\vspace*{-0.4cm}
\end{table}
Table~\ref{table:data_collection} summarizes the data we collected for our experiments. In total, {\tool} downloaded and analyzed 22,330 apps belonging to 26 categories from Google Play Store. For each category, we chose only top-ranked apps in the \emph{``Free''} and \emph{``New Free''} lists. This selection is based on the assumption that the top-ranked apps would have high quality code, and thus, would contain API usages of high interest for learning. We excluded apps in the ``Game'' category because they are often developed using game engines, thus, contain few actual API usages.

Since Android mobile apps are distributed as \code{.apk} files, {\tool} unpacked each downloaded \code{.apk} file and kept only its \code{.dex} file, which contains all bytecode of the app. The total storage space for the \code{.dex} files of all downloaded apps is around 68 GB. {\tool} parsed those \code{.dex} files, which contain nearly 5 million classes. It analyzed each class and looked for methods implemented in that class to build {\model} models. Since an Android mobile app is self-contained, its \code{.dex} file contains bytecode of all external libraries it uses. That leads to the duplication of bytecode of shared libraries. Thus, {\tool} maintains a dictionary of the methods it has already analyzed to help it to analyze each method only once. Because {\tool} built {\model} for all possible execution paths, the number of {\model} could be exponential to the number of branching points in the code. Thus, to make the building process feasible, we excluded methods with more than 10 \code{if} instructions. We also excluded getter and setter methods, which are often short (less than 7 instructions) and do not contain API usages. The remaining consists more than 9 million methods which have in total nearly 260 million bytecode instructions.

\subsection{Extracting API Method Sequences}

\begin{table}[!t]
	\caption{Extracting API method sequences}
	\centering
	\label{table:sequence_extraction}
	\begin{tabular}{lrr}
	\toprule
&  Single object & Multiple object \\
\midrule
	Number of distinct object types (sets) & 2,708  & 24,192 \\
	Total number of method sequences        & 17,973,832 &    20,611,024   \\
	Average number of method sequences           &      6,637   & 852 \\
	Average length of method sequences &   3.8  & 8.0    \\
\bottomrule
\end{tabular}
\vspace*{-0.4cm}
\end{table}

{\tool} built {\model} for each remaining method in the dataset and extracted API method sequences from those models. It extracted sequences for both single object usages and multiple object usages. Since we focus on learning Android API usages, only sequences involving classes and methods of Android application framework were extracted. Sequences with only one method call were disregarded.

Table \ref{table:sequence_extraction} summarizes this sequence extraction process. In total, {\tool} has extracted nearly 18 millions API method sequences involving single object usages of more than 2,700 different API object types. There are more than 24,000 distinct usage-dependent object sets made from those types and {\tool} has extracted over 20 millions method sequences for those sets (i.e. these sequences representing multiple object usages). The running time and space is acceptable: {\tool} took 9 hours to build {\model} and extract those method sequences (i.e. 1.5 seconds per app on average) and needed less than 300 MB of working memory.

\subsection{Training API Usage Models}

\begin{table}[!t]
	\caption{Training API usage models}
	\centering
	\label{table:training}
	\begin{tabular}{lrr}
		\toprule
&  Single object & Multiple object \\
\midrule
		Number of trained models  & 1,152 & 7,228    \\
		Average number of hidden states & 8.1  & 8.3  \\
		Total training time                      & 10 min & 70 min  \\
		Total space to store trained models      & 7.2 MB & 18.5 MB  \\
\bottomrule
\end{tabular}
\vspace*{-0.4cm}
\end{table}
Once all API method sequences are extracted and stored, {\tool} trained {\hmm} models for each object type or usage-dependent object sets. For example, a {\hmm} is trained for the usages involving any \code{MediaRecorder} object and another {\hmm} is trained for the usages involving any two objects \{\code{FileReader}, \code{BufferedReader}\}. We used only 80\% of method sequences associated with each {\hmm} for training (of which 12.5\% was randomly selected as the validation set). The 20\% remaining was held out as testing data for the experiments with API recommendation tasks. To make the training process stable, we did not train {\hmm} models having insufficient method sequences (we used the cut-off threshold of 25 sequences). The training process is summarized in Table~\ref{table:training}. As seen, it is time- and space-efficient: More than 8,000 usage models were trained in about 80 minutes and stored using 26 MB.

\subsection{Sensitivity Analysis}

\begin{figure}[!t]
	\hspace{-15pt}
	\includegraphics[scale = 0.67]{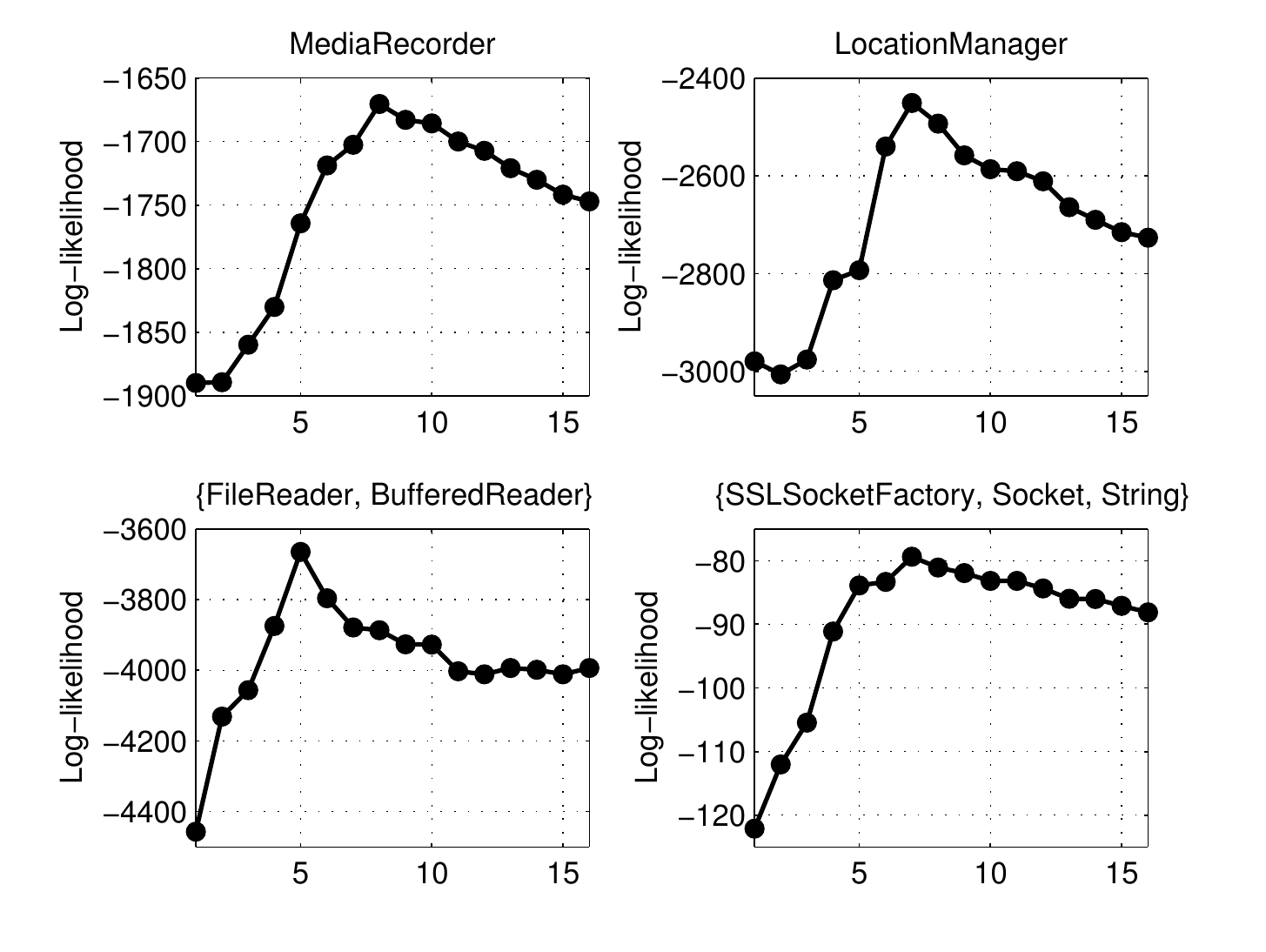}
	\vspace*{-0.8cm}
    \caption{Predictive power of HAPI models with different numbers of internal hidden states}
	\label{number_hiddens}
	\vspace*{-0.8cm}
\end{figure}

Table~\ref{table:training} suggests that {\hmm} models often have 8 internal states. To further study the effect of the number of hidden states $K$ on the predictive power of {\hmm} models, we performed an extra experiment. In this experiment, we trained four {\hmm} models for both single object and multiple objects usages, varied $K$ from 1 to 16 and measured the log-likelihood as their predictive power on validation data.

Figure~\ref{number_hiddens} illustrates the experiment results. As seen, the log-likelihood curves for four {\hmm} models have the same shape. When $K$ is small $(K = 1, 2, 3)$, the predictive power is low. This is reasonable because the models have too few hidden states to distinguish method sequences of different usage patterns. Then, when $K$ increases, the predictive power  increases. However, after reaching a certain point, the predictive power starts to decrease because the hidden states might become redundant and overlap with each other, making the {\hmm} model overfit to the training data and under-performed on the validation data.

The optimal number of hidden states of each model varies and is likely dependent on usage scenarios of the corresponding API objects. For example, as seen in Section II, an \code{MediaRecorder} has many different usage scenarios. However, the object pairs \{\code{FileReader}, \code{BufferedReader}\} are usually used for reading files with a dominant usage scenario illustrated in Figure \ref{hapi_example}. Thus, {\hmm} for \code{MediaRecorder} has more internal states than one for \{\code{FileReader}, \code{BufferedReader}\}.

\subsection{API Usage Recommendation}
In this experiment, we aim to measure the accuracy of {\hmm} models in recommending API usages. Following prior work, we design two different recommendation tasks.

\noindent\textbf{Task 1 - Predicting next call}. An API method sequence is given and the model needs to recommend the most probable next call. This task has been used in~\cite{grapacc, cscc} for the evaluation.

\noindent\textbf{Task 2 - Filling in a hole}. The given API method sequence has a method call missing at a specified position, and the model needs to recommend the most probable call for that position. This task has been used in~\cite{ngram} for evaluation.

We chose $n$-gram model as the baseline for comparison due to several reasons. First, it is a statistical model, thus, is comparable to {\hmm}, which is also a statistical model. In addition, it is widely used in recent research on code completion~\cite{hindle_naturalness, SLAMC}. Most importantly, $n$-gram model for API usages is studied by Raychev et al.~\cite{ngram} and reported as one of the state-of-the-art approaches. 

An $n$-gram model learns all possible conditional probabilities $P(c|S)$, where $c$ is a method call and $S$ is a method sequence of $n-1$ calls. This is the probability $c$ occurs as the next method call of $S$. Using the chaining rule, we use an $n$-gram model to compute the generating probabilities of any given method sequence. In our experiment, we used a 3-gram model, i.e. the occurrence probability of a method call depends on its two previous calls. We trained a 3-gram model for each object type (or set) using the same training data as we trained {\hmm} models. Both models also use the same test data.

\subsubsection{Task 1 - Predicting next call}
\begin{figure}[t]
	\hspace{-20pt}
	\includegraphics[scale = 0.32]{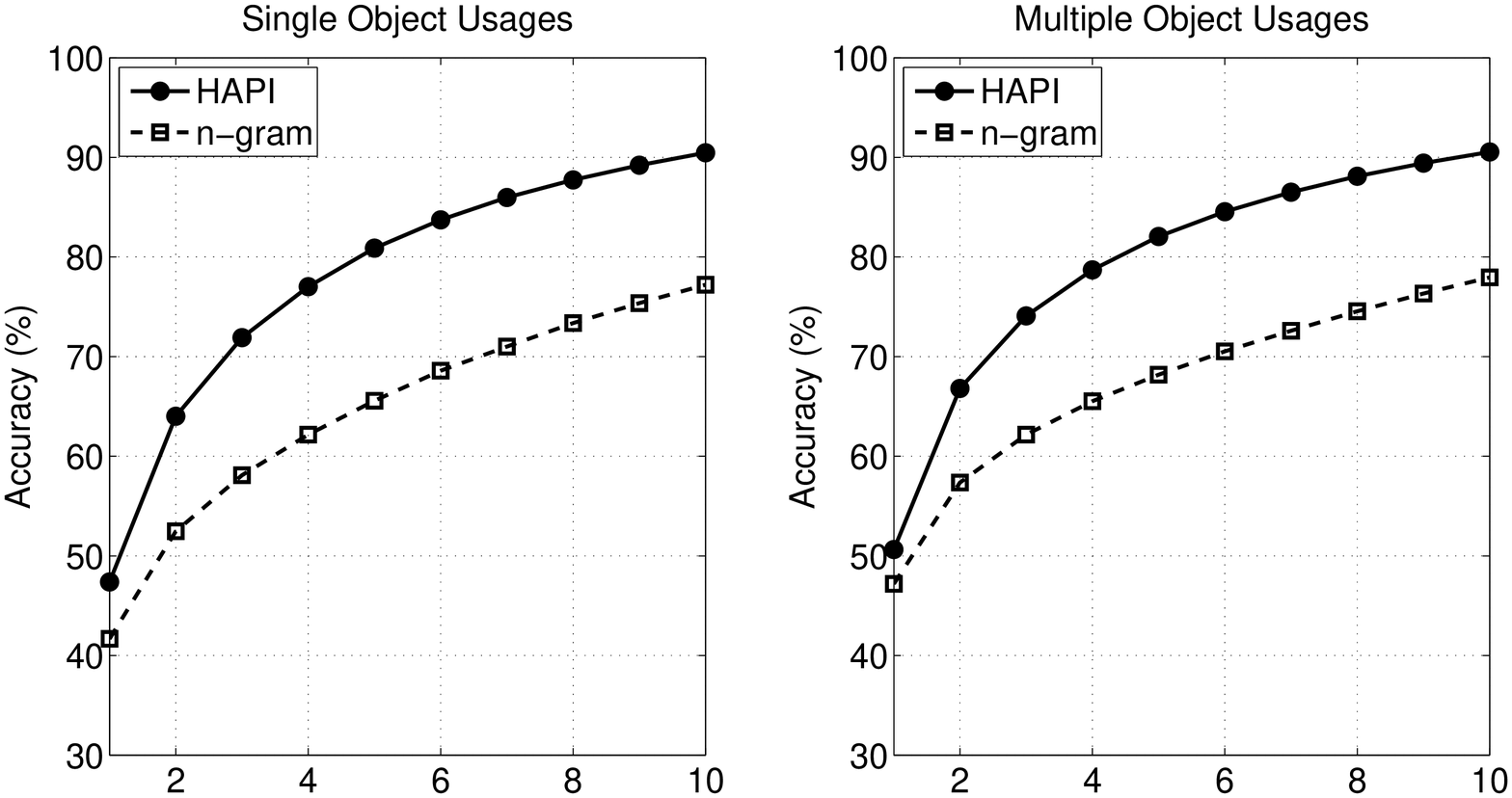}
	\vspace*{-0.6cm}
	\caption{Accuracy on Task 1 - Predicting next call}
	\label{next_result}
	\vspace*{-0.8cm}
\end{figure}

In this task, we evaluated all method calls in every sequence in the testing set. For a method call $c_i$ at position $i$, we used its prefix of $i-1$ method calls as the input of a model and received its recommendation of the top-$k$ most probable next method calls $R_k = \{r_1 , r_2 , ..., r_k\}$. If $c_i$ is in $R_k$, we consider it as an accurate recommendation (i.e. a hit). The top-$k$ accuracy is the ratio of the total hits over the total number of evaluated method calls.

Figure \ref{next_result} shows the experiment results. As seen in the charts, the accuracy is consistent for both single and multiple object usages. More importantly, {\hmm} can recommend API usage for this task with a high level of accuracy. For example, for single object usage, it has a top-3 accuracy of 71.9\% and a top-10 accuracy of around 90.5\%. In addition, {\hmm} always outperforms $n$-gram model. For example, the top-3 and top-10 accuracy results of $n$-gram model for single object usages are 58.1\% and 77.2\%, respectively. The average improvement of {\hmm} over $n$-gram model across all settings is 12.6\%.

\subsubsection{Task 2 - Filling in a hole}

\begin{figure}[t]
	\hspace{-20pt}
	\includegraphics[scale = 0.32]{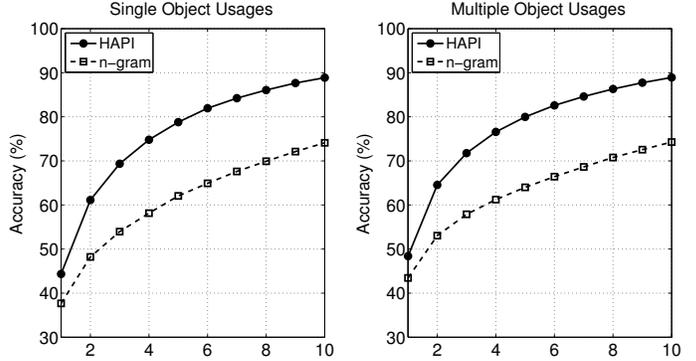}
	\vspace*{-0.6cm}
	\caption{Accuracy on Task 2 - Filling in a hole}
	\label{fill_result}
	\vspace*{-0.8cm}
\end{figure}

In this task, for each method sequence in a test set, we removed a method call at a random position (i.e. making a hole). Then, each available API method was substituted into that position and both {\hmm} and $n$-gram models were used to compute the probability of the entire sequence, respectively. Top-$k$ methods that made the $k$ highest probabilities are selected as the recommendation. If the original method is one among them, it is counted as an accurate recommendation (i.e. a hit). The top-$k$ accuracy is the ratio of the total hits over the total number of evaluated method sequences.

Figure \ref{fill_result} shows the experiment results. As seen in the charts, the accuracy is consistent for both single and multiple object usages and is close to the accuracy for Task 1. The conclusions are also the same as for the experiment for Task 1. That is, i) {\hmm} can recommend API usage for this task with a high level of accuracy and ii) {\hmm} always outperforms $n$-gram model. For example, for single object usages, the top-3 and top-10 accuracy results of {\hmm} are 69.4\% and 88.9\%, while those of $n$-gram model are 54\% and 74.1\%, respectively. The average improvement of {\hmm} over $n$-gram model across all settings is 14.4\%.

The results suggests that for ``Filling in the hole'' task, {\hmm} model has higher improvement over $n$-gram model than for ``Predicting next call'' task. One explanation for this result is that in ``Filling in the hole'' task, {\hmm} has the whole sequence as the context for prediction, while in ``Predicting next call'' task, it has only the first part of the sequence. In both tasks, $n$-gram always uses only two prior calls.

\section{Related Work}
There exist several works that proposed statistical models for learning API usages. 
Raychev et al. \cite{ngram} use $n$-gram and recurrent neural networks (RNN) to learn API usage patterns per object which are used to predict and suggest next API calls. 
Nguyen et al. \cite{gralan} presented GraLan, a graph-based statistical language model that learns common API usage (sub)graphs from a source code corpus and computes the probabilities of generating new usage graphs given the observed (sub)graphs. 

Statistical language models have been successfully used to capture patterns in source code. 
Hindle et al. \cite{hindle_naturalness} showes that source code is repetitive and predictable like natural language and they adopted $n$-gram model on lexical tokens to suggest the next token. 
SLAMC \cite{SLAMC} represents code by semantic tokens, i.e. annotations of data types, method/field signatures, etc. rather than lexical tokens. SLAMC combines $n$-gram modeling of consecutive semantic tokens, topic modeling of the whole code corpus, and bi-gram of related API functions.
Tu et al. \cite{tu_localness} exploited the localness of source code and improve $n$-gram language model with caching for recently seen code tokens to improve next-token suggestion accuracy  
Allamanis and Sutton \cite{allamanis_massive} trains $n$-gram language model a giga-token source code corpus. 
NATURALIZE \cite{NATURALIZE} use $n$-gram language model to learns the style of a codebase and suggest natural identifier names and formatting conventions.
Jacob et al. \cite{jacob_template} uses $n$-gram model to learn code templates.
Hidden Markov Model has been used infer the next token from user-provided abbreviations \cite{hmm_abbreviated} and detect coded information islands, such as source code, stack traces, and patches, from free tex \cite{hmm_islands}.
Maddison et al. \cite{maddison_tree} proposed tree-based generative models for source code. 
Hsiao et al. \cite{ngram_web} learns $n$-gram language model on program dependence graph and uses the model for finding plagiarized code pairs.

Pattern mining approaches represent usage patterns using various data structure such as sequences, sets, trees, and graphs.
JADET \cite{jadet} extracted a usage model in term of a finite state automaton (FSA). 
MAPO \cite{mapo} mined frequent API call sequences and suggests associated code examples. 
Wang et al. \cite{upminer} mines succinct and high-coverage API usage patterns from source code.
Acharya et al. \cite{taoxie_usage_scenarios} proposed an approach to mine partial orders among APIs.
Buse and Weimer \cite{api_synthesize} propose an automatic technique for mining synthesizing succinct and representative human-readable API examples.
Other mining techniques includes mining associate rules \cite{lo_mining}, item sets \cite{bruch}, subgraphs \cite{groum}, \cite{chang_graphs}, code idioms \cite{idiom} etc.

One application of usage patterns mined from existing code is to support code completion. 
Grapacc \cite{grapacc} mines and stores API usage patterns as graphs and suggest these graphs in current editing code.
Bruch et al. proposed three approaches for code completion. First, FreqCCS recommends the most frequently used method call. Second, ArCCS mines associate rules and suggest methods that often occur together. Finally, a best matching neighbors code completion technique that makes used $k$-nearest-neighbor algorithm.
SLANG \cite{ngram} uses $n$-gram to suggest the next API call based on a window of $n-1$ previous methods.
Precise \cite{parameter_mining} mines existing code bases and builds a parameter usage database. Upon request, it queries the database and recommends API parameters. 
Graphite \cite{active_completion} allows library developers to introduce interactive and highly-specialized code generation interfaces that could interact with users and generates appropriate source code.

Other approaches have been proposed to improve code completion tasks.  
Robbes et al.\cite{history_completion} improves code completion with program history. They measure the accuracy of replaying entire change history of programs with completion engine and gather information to improve the engine. 
Hou and Pletcher \cite{sorting_api} found that ranking method calls by frequency of past use is effective and propose new strategies for organizing APIs in the code completion pop up.
Hill and Rideout \cite{auto_completion} matches the fragment under editing with small similar-structure code segments that frequently exist in large software projects.
McMillan et al. \cite{api_doc}  and Subramanian et al. \cite{api_live} use API documentation to suggest source code examples to developers.
Holmes and Murphy \cite{structural_source_code} describe an approach for locating relevant code examples based on heuristically matching with the structure of the code under editing.

\section{Conclusion}
We propose a statistical approach to learn API usages from bytecode of Android mobile apps. We propose a graph-based representation of usage scenarios to extract API method sequences from bytecode and use those sequences to learn API usage models based on Hidden Markov Models. Our empirical evaluation on a large dataset indicates that our approach can learn useful API usage models which can provide higher levels of accuracy in API recommendation than the $n$-gram model.






\bibliographystyle{IEEEtran}
\bibliography{androidpm}
%
%
%
\end{document}